\documentclass[a4paper]{article}
\usepackage{graphicx}
\usepackage[top=3.0cm,bottom=3.0cm,left=3.0cm,right=2.0cm]{geometry}
\usepackage{amssymb,amsmath}
\usepackage{textcomp}
\usepackage{subfigure}
\usepackage{cite}
\usepackage{braket}
\usepackage{cleveref}
\usepackage{slashed}
\usepackage{stackrel}
\usepackage[0]{editing}
\usepackage{kantlipsum}
\usepackage{sidecap}
\usepackage{ulem} 

\usepackage[shortcuts,acronym,nonumberlist,nomain]{glossaries}

\usepackage{babel}
\title{Nambu--Jona-Lasinio model with a fractal inspired coupling}
\author{E. Meg\'{\i}as$^{1}$, M. J. Teixeira$^{2, 3}$, V. S. Tim\'oteo$^{3}$ and A. Deppman$^{2}$}

	\date{
	\begin{small}
		$^1$ Departamento de F\'{\i}sica At\'omica, Molecular y Nuclear and Instituto Carlos I de F\'{\i}sica Te\'orica y Computacional, Universidad de Granada, 18071 Granada, Spain \\
		$^2$ Instituto de F\'isica, Universidade de S\~ao Paulo, 05508-090 S\~ao Paulo, SP, Brazil \\
		$^3$ Grupo de \'Optica e Modelagem Num\'erica, Faculdade de Tecnologia - GOMNI/FT \\ Universidade Estadual de Campinas - UNICAMP, 13484-332 Limeira, SP, Brazil \\
	\end{small}	
	\vspace{0.5cm}
	\today
	}

\newacronym{bcs}{BCS}{Bardeen, Cooper and Schrieffer}
\newacronym{njl}{NJL}{Nambu-Jona-Lasinio}
\newacronym{kg}{KG}{Klein-Gordon}
\newacronym{qcd}{QCD}{Quantum Chromodynamics}
\newacronym{qft}{QFT}{Quantum Field Theory}
\newacronym{tf}{TF}{thermofractal}
\newacronym{qgp}{QGP}{Quark-Gluon Plasma}
\newacronym{gor}{GOR}{Gell-Mann-Oakes-Renner}
\newacronym{hep}{HEP}{High Energy Physics}
\newacronym{hrg}{HRG}{Hadron Resonance Gas}
\newacronym{fnjl}{fNJL}{fractal Nambu-Jona-Lasinio}
\newacronym{sct}{SCT}{Self-Consistent Thermodynamics}
\newacronym{bst}{BST}{Bootstrap Model}

\begin{document}

\maketitle

\bigskipamount=1cm

\begin{abstract}
\noindent
The Nambu--Jona-Lasino model is modified by the inclusion of a running-coupling that was obtained by a fractal approach to Quantum Chromodynamics. The coupling follows a $q$-exponential function and, in the context of high energy collisions, explains the origin of the Tsallis non-extensive statistics distributions. The parameter $q$ is completely determined in terms of the number of colours and the number of quark flavours. We study several aspects of the extended model and compare our results to the standard NJL model, where a constant coupling is used in combination with a sharp cutoff to regularize the gap equation. We show that the modified coupling regularizes the model in a smooth cutoff fashion and reproduces the pion mass and decay constant, providing an almost identical Gell-Mann-Oakes-Renner relation as in the standard NJL model. In both models the relation is satisfied in similar cutoff scales. An important novelty of this work is the physical explanation, in terms of the fractal QCD vacuum, for a running coupling that renormalizes the quark condensate.
\end{abstract}


\section{Introduction}

Before the quark structure of the hadrons was discovered, many phenomenological models using hadronic degrees of freedom appeared. Some of them were so successful in explaining new aspects of the strongly interacting systems that they are studied up to these days. The \gls{hrg}~\cite{Hagedorn:1984hz,Agasian:2001bj,Megias:2012kb} and the \gls{njl}~\cite{Nambu:1961tp,Nambu:1961fr} models are examples of those phenomenological models.

The \gls{njl} model was proposed as a way to describe the nucleon self-energy interaction in a self-consistent way. The model allowed the calculation of the dynamical mass, and exposed a non-trivial solution for the ground-state of the hadron which was not apparent in the perturbative approach \cite{Nambu:1961tp,Nambu:1961fr}. This solution is associated with the formation of a condensate state of the pair of fermion-antifermion, in a process similar to the \gls{bcs} pairing. With the NJL model, it was possible to investigate the effects of the chiral symmetry breaking and to identify the pion as the Goldstone boson \cite{Vogl:1991qt,Klevansky:1992qe,Buballa:2003qv}. Among the drawbacks is the fact that the theory is not renormalizable, so the dependence on the regularization procedure is important.

The model has gained attention recently because of the possibility to study the confined/deconfined regimes of the hadronic matter \cite{Fukushima:2003fw,Ratti:2005jh,Megias:2004hj,Megias:2006bn,Roessner:2006xn}. As the Quark-Gluon Plasma (QGP) reaches the hadronization point, there must be a reduction in the degrees of freedom of the system. New formulations of the \gls{njl} model include the quark degrees of freedom, approximating the phenomenological model to the \gls{qcd} at the non-relativistic regime. \gls{hep} experiments provided much information about the strongly interacting systems. The Higgs bosons were identified, confirming the prediction of the dynamical mechanism for mass formation, and the existence of a deconfined regime of the quark matter, the QGP, was established. The hadronization of the QGP happens at relatively low energy scales, posing important challenges for the perturbative QCD approach. At high momentum processes, the asymptotic freedom allows an accurate use of the perturbative calculations to a few orders, but this is not the case for the transition from confined to deconfined regimes.

The studies on properties of compact stars and the experiments with heavy-ion collisions
have indicated the possibility of having very intense magnetic fields affecting the quark matter. This idea inspired some modifications of the standard NJL model in order to access
the effects of strong magnetic fields in quark matter \cite{Ferreira:2013tba,Farias:2015eea,Pagura:2016pwr,Farias:2016gmy} and on the pion properties \cite{Avancini:2016fgq,Coppola:2018vkw}. Deconfinement and chiral phase transition in quark matter under strong electric field \cite{Tavares:2019mvq}, 
the magnetic field-dependence of the neutral pion mass in the linear sigma model coupled to quarks \cite{Ayala:2018zat} and the anisotropy in the equation of state of strongly magnetized quark matter \cite{Avancini:2017gck} were also studied recently with effective models.

Asymptotic freedom is a consequence of the scaling properties of the \gls{qcd} fields and vertices, as expressed by the so-called Callan-Symanzik \cite{callan1970,symanzik1970} equations. The scaling properties stem from the more basic scaling properties of the Yang-Mills fields, and they are responsible for the possibility to regularize the field theory by imposing cuts at low and high energies, while re-normalizing the coupling, the masses and the fields. The perturbative method in \gls{qft} introduces the concept of effective parton, which carries the effects of self-energy interaction. While this method simplifies the calculations, it endows the parton with a complex structure, and scaling properties. So this non-Abelian field holds the necessary characteristics of a fractal system, and it can be shown that a structure similar to the so-called \glspl{tf}  \cite{Deppman:2016fxs,deppman2018} can be formed in systems with a dynamical evolution regulated by the Yang-Mills field theory. The concept of fractal structures in hadronic systems is closely connected to early \gls{hrg} approaches based on the Hagedorn's \gls{sct}~\cite{hagedorn1965} or on the \gls{bst} proposed by Chew, Frautschi and others~\cite{chew1968,frautschi1971}.

The \glspl{tf} are thermodynamic systems with a fractal structure in their thermodynamic distributions. It was shown that they follow from the non-extensive thermodynamics proposed by Tsallis \cite{Tsallis:1987eu}. The  \glspl{tf} and the Tsallis statistics are intertwined in a deeper way: the algebra of the \gls{tf} transformations group  and the $q$-algebra associated to the Tsallis non-additive entropy are isomorphic. The \gls{tf} structure of the \gls{qft} reconciles the quark-gluon structure with the early attempts to describe the high energy collisions by using self-consistent arguments. The \gls{bst} and the self-consistent thermodynamics by Hagedorn \cite{hagedorn1965} use scaling properties to obtain recurrent relations. Hagedorn's theory can be generalized to a non-extensive self-consistent thermodynamics~\cite{Deppman-2012} that can describe the heavy-tail distributions from the high energy collisions, but also provides a hadron-mass spectrum formula that can describe the observed hadronic states up to masses as low as the pion mass. This result is better than that obtained by the Hagedorn formula.

Tsallis statistics seems to be related to systems with small number of degrees of freedom~\cite{Biro:2014fna,Biro:2012bka}. In the mathematical aspects of Statistical Mechanics, this means that the dimension of the phase space is limited, and $q$ is related to that number of dimensions~\cite{Deppman:2021bov}. This holds for thermofractals but also for an ideal gas with limited number of particles. In this case, as the number of particles increases, also the phase space dimension increases because there is a fixed number of degrees of freedom associated with each particle added. According to this reasoning, any system that keeps the phase space dimension limited to a finite value in the thermodynamical limit would follow the Tsallis statistics, and the reason can be traced down to the topology of the phase space~\cite{Megias:2022yrw}.

The \gls{tf} structure of the Yang-Mills fields leads to an analytic expression for the running coupling constant which is given by
\begin{equation}
	g(\varepsilon_o)=\prod_{i=1}^2\left[1+(q-1)\frac{\varepsilon_i}{\lambda} \right]^{-\frac{1}{q-1}}\,,  \label{eq:ge0}
\end{equation}
with $\varepsilon>0$ and $q>1$, where $\lambda$ is a scale parameter, $q$ is a parameter associated to the number of degrees of freedom, and $\varepsilon_i$ is the energy of the $i$th parton in the interaction. Due to the \gls{tf} algebra we have $\varepsilon_1+\varepsilon_2=\varepsilon_o$, where $\varepsilon_o$ is the total energy of the interacting system. The running-coupling given in Eq.~(\ref{eq:ge0}) results from the fact that the partonic momentum distributions are defined at the vertices.

The parameter $q$, which in the non-extensive statistics is a measure of the non-additivity, can be calculated in the fractal approach to \gls{qft} as a function of the field theory parameters. For \gls{qcd}, in particular, those parameters are the number of colors and the number of flavors. The dependence of the entropic index $q$ with $N_c$ and $N_f$ has been obtained in~\cite{Deppman:2019yno} within the thermofractal approach, leading to
\begin{equation}
    q = 1 + \frac{3}{11 N_c - 2 N_f} \,. \label{eq:q_Nc_Nf}
\end{equation}
The theoretical value of the entropic index is $q = 1 + 1/7 \simeq 1.143$ for 
$(N_c = 3, N_f = 6)$, which is in good agreement with the experimental findings that give
$q=1.14 \pm 0.01$. The coupling of Eq.~(\ref{eq:ge0}) explains why the Tsallis distribution frequently appears in high-energy multi-particle production. In the present work,
we will adopt the value $q = 1 + 3/29 \simeq 1.103$ for $(N_c = 3, N_f = 2)$, since we are 
restricted to the SU(2) version of the model as we make use of the Gell-Mann-Oakes-Renner (GOR) relation~\cite{Gell-Mann:1968hlm}, which was proved for the low-energy limit
using the SU(2) symmetry. 

Although the theoretical value for $q$ has been verified in the context of the deconfined regime of the hadron matter, the result represented by Eq.~(\ref{eq:q_Nc_Nf}) is a consequence of the renormalization group equation, and is valid for QCD everywhere. Some investigations for the case of particle production in the confined regime~\cite{Baptista2024} suggest that the value $q=1.14$ remains valid below the critical point, with the differences in the cross-section being caused by the number of vertices, which changes to accomplish the confinement constraints.

The success in describing the hadron spectrum allows us to conjecture that the same scaling properties of the system formed at high energy collisions is found in the hadronic structure. A consequence is that the running coupling constant described above can be valid in the  non-perturbative processes taking place inside the hadrons. In this work we assume that the partonic interaction is described by the fractal running coupling constant. 

We note that, once the Tsallis distributions emerge from the coupling and with the parameter $q$ with a value already in agreement with that obtained in experimental data analyses, according to the thermofractal approach described in Ref.~\cite{Deppman:2016fxs}, there is no room for other sources of non-extensivity in the system, and a thermal source of non-extensivity must be excluded. In this work, we assume that the nonextensive behaviour arises purely from the running-coupling form, excluding thermodynamical effects.

\section{Standard and fractal NJL models}
\label{sec:quark_condensate}

In this section, we briefly review the basic concepts of the \gls{njl} model that are relevant in this work, and then we introduce the extension of the model by incorporating a fractal inspired coupling. We will refer to this extended model as the \gls{fnjl} model.

\subsection{Brief review of the NJL model}

The \gls{njl} Lagrangian density is given by
\begin{equation}
	\label{eq:lag-njl}
	\mathcal{L}_{\mathrm{NJL}} = - \bar{\psi} \left( i \gamma^\mu \partial_\mu  - m_0\right) \psi + G\left[  \left( \bar{\psi} \psi\right) ^2  - \left( \bar{\psi} i \gamma_5\psi\right) ^2\right]. 
\end{equation}
The second term at the right-hand side of the Lagrangian density incorporates the self-energy interaction. The method proposed by Nambu and Jona-Lasino allows to obtain the self-energy contribution in a self-consistent way, and leads to the dynamical fermion mass given by the gap equation
\begin{equation}
	m = m_0 -2 N_f ~ G ~ \langle \bar\psi \psi\rangle \,, \label{gapequation}
\end{equation}
where $G$ is constant and $\langle \bar\psi \psi \rangle$ is the quark condensate. In the mean field approximation, this condensate is given by~\cite{Vogl:1991qt}
\begin{equation}
    \langle \bar\psi \psi \rangle = -i \, \textrm{tr} S_F(0) \,, 
    \label{eq:cond}
\end{equation}
with the Dirac propagator 
\begin{equation}
\label{eq:dirprop}
    S_F(x-y)  = \int \frac{d^4p}{(2\pi)^4} \frac{e^{ip \cdot (x-y)}}{\slash{\!\!\!p} - m + i \epsilon} \,.
\end{equation}

The physical meaning of Eq.~\eqref{eq:cond} is the interaction between a single fermion with
the particles at Dirac sea with momenta $\vec{p}$. The integral in Eq.~\eqref{eq:dirprop} is
divergent and a three momentum cutoff $\Lambda$ is usually introduced to regularize the gap
equation \cite{Vogl:1991qt}. In this case, after integration in~$p^0$ the resulting integral is
\begin{equation}
\langle \bar\psi \psi \rangle = -\frac{N_c}{\pi^2} \int_0^\Lambda dp \, p^2 \frac{m}{E_p}  \,, 
\end{equation}
where $p \equiv |\vec{p}|$, and $E_p = \sqrt{p^2 + m^2}$ is the particle energy. The gap equation in the 
NJL model is then 
\begin{equation}
m = m_0 + \frac{4 m N_c N_f}{2\pi^2} \int_0^\Lambda dp \, p^2 \frac{G}{\sqrt{p^2 + m^2}} \, . \label{eq:gap_NJL}
\end{equation}
In the chiral symmetric regime, $m_0=0$, we get the self-consistent relation
\begin{equation}
 1 = \frac{4 N_c N_f}{2\pi^2} G  \int_0^{\Lambda} dp \, p^2 \frac{1}{\sqrt{p^2 + m^2}} \,. \label{eq:gap_m0}
\end{equation}
There is a critical value of the coupling $G$ below which the quark condensate and dynamical mass vanish: $\langle \bar\psi \psi \rangle = 0$ and $m=0$. From Eq.~(\ref{eq:gap_m0}), this critical value turns out to be
\begin{equation}
G_{\textrm{crit}} = \frac{\pi^2}{N_c N_f} \frac{1}{\Lambda^2} \,. \label{eq:GcNJL}
\end{equation}
When the coupling constant is greater than $G_{\textrm{crit}}$, the vacuum is in a state with a non-vanishing condensate given by Eq.~(\ref{eq:cond}), so that $\langle \bar\psi \psi\rangle$ is considered as the order parameter for the phase transition between the
Wigner-Weyl and Nambu-Goldstone phases. Within the \gls{njl} model, the result of the chiral condensate at large values of the coupling turns out to be
\begin{equation}
\langle \bar\psi \psi \rangle \stackrel[G \gg 1/\Lambda^2]{\simeq}{} -\frac{N_c}{3\pi^2} \Lambda^3 + \frac{9 \pi^2}{40 N_c N_f^2 \Lambda} \frac{1}{G^2} \;.  \label{eq:qqNJL_largeG}
\end{equation}

The GOR relation \cite{Gell-Mann:1968hlm} is a model independent result obtained by considering the SU(2) symmetry group effects on the low-energy domain of the hadron structure. The relation represents a sum-rule that relates the pion mass, $m_{\pi}$, and the pion decay constant, $f_{\pi}$, to the current quark mass, $m_0$, and the quark condensate (scalar density). This relation is
\begin{equation}
    m_\pi^2 f_\pi^2 = -N_f m_0 \langle \bar \psi \psi \rangle + {\mathcal O}(m_0^2 f_\pi^2) \,.  \label{eq:GOR}
\end{equation}
By using
\begin{equation}
    m_\pi = 140 \, \textrm{MeV} \,, \qquad f_\pi = 92 \, \textrm{MeV} \,, \qquad m_0 = 5 \, \textrm{MeV} \,,
\end{equation}
one gets the physical value of the chiral condensate
\begin{equation}
    \langle \bar\psi \psi \rangle_{\textrm{Phys}} = -(0.255 \, \textrm{GeV})^3 \,. \label{eq:qq_physical}
\end{equation}
With this value, one gets for the \gls{njl} model
\begin{equation}
    G_{\textrm{Phys}} = 5.22 \, \textrm{GeV}^{-2} \,, \qquad \textrm{for} \qquad \Lambda = 0.65 \, \textrm{GeV} \,.
\end{equation}

\begin{figure}[t]
	\centering
	\includegraphics[width=0.35\linewidth]{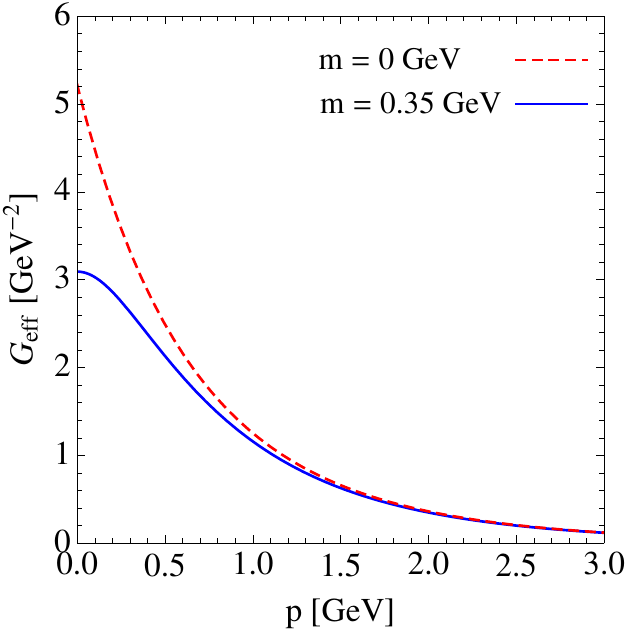} 
	\caption{Effective two-body coupling as a function of $p$, cf. Eq.~(\ref{eq:Geff}). We display the result for zero (dashed red) and finite (solid blue) dynamical quark masses. We have used the values $G_q = 5.22 \, \textrm{GeV}^{-2}$, $\lambda = 0.65 \, \textrm{GeV}$ and $q = 1.103$.}
	\label{fig:Geff}
\end{figure}
%

\subsection{The fractal NJL model} 
In order to introduce the fractal structure that emerges from the scaling properties associated to the renormalization of QCD, we implement a running coupling of Eq.~(\ref{eq:ge0}) which is a direct consequence of the fractal structure~\cite{Deppman:2016fxs,deppman2018}. Therefore, the effects of the fractal structure can be implemented by substituting the constant coupling $G$ of the \gls{njl} model by the fractal running-coupling in the gap equation.

If we consider that one of the particles has zero energy, the effective two-body
coupling can be written as
\begin{equation}
    G_{\rm eff} (p~;~G_q,~\lambda~,~q) = G_q  \cdot
    \left[ 1 + (q-1)\frac{\sqrt{p^2+m^2}}{\lambda} \right]^{-\frac{1}{q-1}} \;,  \label{eq:Geff}
\end{equation}
where we identify $G_q$ as the strength of the coupling. The behavior of $G_{\rm eff}(p)$ is displayed in Fig.~\ref{fig:Geff}, where we observe that
the coupling decreases and approaches zero asymptotically. This is in clear contrast
with the constant coupling, $G$, of the standard \gls{njl} model. In  Fig.~\ref{fig:Geff} we
display the effective coupling in two cases: the chiral limit ($m=0$) and with broken chiral
symmetry using $m=0.35$~GeV. In both cases, the effective coupling asymptotically approaches zero when
the momentum goes to infinity. Then, the effective coupling eliminates the divergences and regularizes the fractal \gls{njl} model.

After the introduction of the running-coupling, the gap equation then reads
\begin{equation}
m = m_0 + \frac{4 m N_c N_f}{2\pi^2} \int_0^\infty dp \, p^2 \frac{G_{\rm eff}(p)}{\sqrt{p^2 + m^2}}   \; , \label{eq:F}
\end{equation}
for the case where $q>1$. The parameter $\lambda$ in the $q$-exponential function of Eq.~(\ref{eq:Geff}) plays the role of a smooth cut-off and the quark condensate may be computed as $-(m - m_0)/(2 N_f G_q)$. Notice that the integrand of Eq.~(\ref{eq:F}) behaves as $\propto p^{1-\frac{1}{q-1}}$ at large momentum, so that the integral is convergent in the UV only if $q < 3/2$. For the case $q<1$, the integration in momentum should be performed in the finite interval
 \begin{equation}
 0 < p < \sqrt{\left(\frac{\lambda}{1-q} \right)^2-m^2} \;. \label{eq:p_interval}
 \end{equation} 
 The upper limit corresponds to the momentum at which the $q$-exponential function vanishes. Then the integral in Eq.~\eqref{eq:F} is convergent also for $q<1$.

 While the coupling of Eq.~(\ref{eq:Geff}) plays the role of a regulator, it is important to stress at this point that beyond phenomenological considerations, this regulator has a physical motivation based on the thermofractal approach to QCD. Let us mention that other regulators have been introduced in the literature that are connected to non-local interactions~\cite{Bowler:1994ir,Praszalowicz:2002nc,GomezDumm:2001fz,Radzhabov:2004ub,Hell:2008cc,Contrera:2009hk}, as well as proper time regularization prescriptions, which allow the study of chaotic dynamics~\cite{Ahmad:2018grh}.
 
In the \gls{fnjl} model, the critical coupling in the chiral symmetric regime, $m_0 = 0$, is
\begin{equation}
 \qquad G_{q,{\textrm{crit}}}  = \frac{\pi^2}{2N_c N_f} \frac{(2-q)(3-2q)}{\lambda^2}  \,, \qquad ( 0 < q < 3/2) \,. \label{eq:Gc}
\end{equation}
For the sake of comparison, in the following we will compare the behaviour of some quantities in the \gls{njl} and in the f\gls{njl} models. For this comparison, we choose the parameter $\lambda$ in such a way that the critical point for both models coincide, that is,
$G_{\textrm{crit}} = G_{q,\textrm{crit}}$. This leads to the following relation between the cut-offs
\begin{equation}
    \lambda = \Lambda \sqrt{\frac{(2-q)(3-2q)}{2}} \,. \label{eq:lambda_Lambda}
\end{equation}
Notice that one has $\lambda = \Lambda/\sqrt{2}$ for $q = 1$. More generally Eq.~(\ref{eq:lambda_Lambda}) implies $\lambda < \Lambda$ for $0.72 < q < 3/2$.

\begin{figure}
	\centering
     \includegraphics[width=0.4\linewidth]{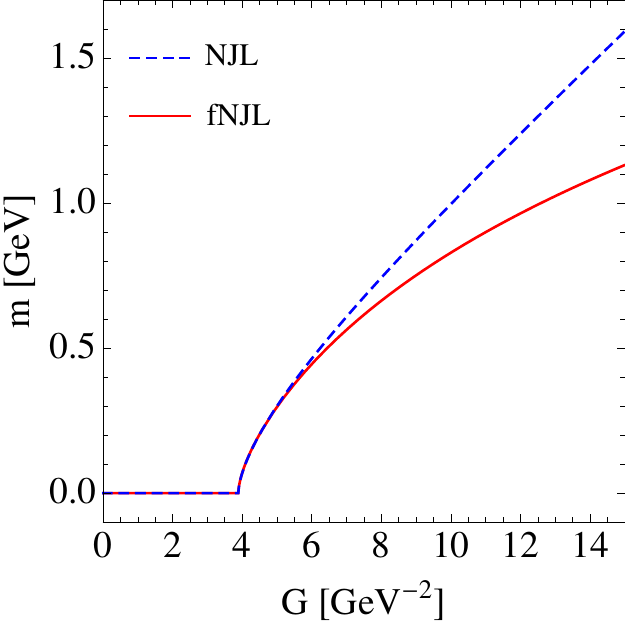} \hspace{0.5cm}
	\includegraphics[width=0.43\linewidth]{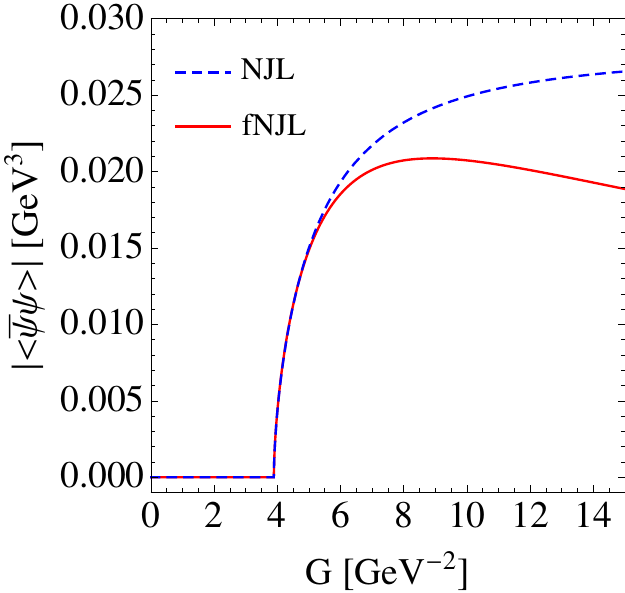}
	\caption{Gap mass (left panel) and condensate (right panel) as a function of the coupling. The critical value of $G$ is the same in both models: $G_{\textrm{crit}} = 3.89 \, \textrm{GeV}^{-2}$. This is obtained with $\Lambda = 0.65 \, \textrm{GeV}$ and $\lambda = 0.388 \, \textrm{GeV}$. Here we have considered $q = 1.103$ and $m_0 = 0$. Note that both models match up to $G \approx 6~{\rm GeV}^{-2}$. The label $G$ refers either to the coupling $G$ for the NJL model, or $G_q$ for the \gls{fnjl} model.}
	\label{fig:plotLLq2}
\end{figure}

With this choice of parameters, we obtain the results in Fig.~\ref{fig:plotLLq2}.
We observe that the behaviour of the mass gap, $m$, and the condensate $\langle \bar \psi \psi \rangle$ for both models are similar near the critical point for the choice of parameters we used. Therefore, despite that the two models have been defined by different effective couplings $G_{\rm eff} (p)$, the results are similar in a wide range of the couplings $G$ and $G_q$. The physical value for the pion mass is obtained by both the \gls{njl} and the f\gls{njl} models for similar values of $G$ and $G_q$. This result is obtained because the effective coupling of the f\gls{njl} model decreases fast near the value of the cut-off momentum.

It is important to stress that the use of the fractal inspired coupling leads to a $q$-exponential running coupling which renormalizes the gap equation. In Ref.~\cite{Rozynek:2015zca}, a q\gls{njl} model was proposed through a statistical approach, where the Tsallis entropy~\cite{Tsallis:1987eu} was used in Eq.~(\ref{eq:cond}) to extend the \gls{njl} model. This model was extended to include the Polyakov loop (qPNJL) in Ref.~\cite{Zhao:2020xob}. Other related works include Ref.~\cite{Mitra:2017cib}, where it was introduced  an effective coupling from a transport approach by calculating the thermal mass from semi-classical transport theory inspired by non-additive statistics.  In Ref.~\cite{Islam:2023zpl}, the effective mass was computed from another type of qNJL model and its variation was studied along with thermal and magnetic effects and different values of the entropic index $q$. It would be interesting to study the similarities between the q\gls{njl} models studied in these references and the \gls{fnjl} model proposed here. On the one hand, let us stress that in these references the parameter $\lambda$ turns out to be connected to the temperature. However, the fundamental point of the present work is to show how a renormalized interaction can emerge from a fractal QCD vacuum, an approach that leads to a nonextensive version of the \gls{njl} model as well.  On the other hand, in the present work we will follow a different path. Since our approach allows us to connect the entropic index, $q$, to the fundamental parameters of the QFT, as the number of flavours and colours in \gls{qcd}, we will investigate the behaviour of the f\gls{njl} model when the parameter $q$ changes, and then we will associate this to different QFTs, where the number of flavours and colours may differ from those of the \gls{qcd}. A more detailed comparison between our fractal model and those proposed in Refs.~\cite{Rozynek:2015zca,Zhao:2020xob,Mitra:2017cib,Islam:2023zpl} will be carried out soon.

Our fractal inspired approach may also be extended to the three-flavour 
NJL model under the assumption that all interactions handled in the mean field
approximation are affected in the same way. In this case, we would have
two effective running couplings in the three-flavour NJL model. We will 
present the fractal three-flavour model in a forthcoming work.

\subsection{Behaviour of the fractal NJL model}

The f\gls{njl} model will present some advantages if it can go beyond the standard \gls{njl} model. To verify how it behaves in different conditions, we now look for the parameters that reproduce the values of some important physical quantities. Here, we use the pions mass and the GOR relation as the physical quantities to fix the parameters of the f\gls{njl} model. The assumption that the physical values of $m_\pi$ and $f_\pi$ are reproduced in both models, does not lead necessarily to the same critical values for the couplings. In fact, by using the expressions of Eqs.~(\ref{eq:GcNJL}) and~(\ref{eq:Gc}) we get 
\begin{equation}
    \frac{G_{q,\textrm{crit}}}{G_{\textrm{crit}}} = \frac{1}{2} (2-q)(3-2q) \left( \frac{\Lambda}{\lambda} \right)^2  \,.
\end{equation}
This formula implies that $G_{q,{\textrm{crit}}} < G_{{\textrm{crit}}}$ for  $\lambda \simeq \Lambda$ and $0.72 < q < 3/2$.

In the pursuit for the values of the f\gls{njl} parameters that allow to reproduce the physical quantities, we use the \gls{njl} model as a guide. We calculate the condensate density using the \gls{njl} model with the parameters with which the model reproduces the physical values for the pion mass and its decay width, and then find the best values for $G_q$, $\lambda$ and $q$ that reproduce the same condensate density.

The condensate density can be calculated analytically in the asymptotic regime $m \gg \lambda$ for any values of $0<q<3/2$. The asymptotic expression is
\begin{equation}
    \langle \bar\psi \psi \rangle \stackbin[G_q \gg 1/\lambda^2]{\simeq}{} \left\{ 
\begin{array}{cl}    
    & - 2^{-\frac{2-q}{3-2q}} \pi^{-\frac{3}{2}\frac{q-1}{3 - 2q}} N_c^{\frac{q-1}{3 - 2q}} (N_f G_q)^{\frac{3q-4}{3-2q}} \left( \frac{\lambda}{q-1} \right)^{\frac{1}{3-2q}}  \left( \frac{\Gamma\left[ \frac{3-2q}{2(q-1)} \right]}{\Gamma\left[ \frac{q}{2(q-1)}\right]}\right)^{\frac{q-1}{3-2q}} \,; \qquad ~~~(1 < q < 3/2)  \\
     & -\frac{\lambda}{2 N_f G_q} \log\left(  \frac{N_c N_f}{\pi^{3/2}} G_q \lambda^2 \right)   \,;  \qquad \qquad ~~~~~~~~~~~~~~~~~~~~~~~  \qquad \qquad \qquad \qquad (q=1) \\
    & \frac{1}{2 N_f G_q} \left[ -\left(\frac{\lambda}{1-q}\right) + \left( \frac{\pi^3}{2 N_c^2 N_f^2 G_q^2}\right)^{\frac{1-q}{5-3q}} \left( \frac{\lambda}{1-q} \right)^{\frac{1+q}{5-3q}} \left( \frac{\Gamma\left[ \frac{7-5q}{2(1-q)} \right]}{\Gamma\left[ \frac{2-q}{1-q}\right]}\right)^{\frac{2(1-q)}{5-3q}}  \right] \,;  \qquad  (q < 1)
    \end{array} \,. \right.  \label{eq:qq_asymptotic}
\end{equation}
At large value of the coupling the chiral condensate presents a power-law behaviour $\langle \bar\psi \psi \rangle \propto G_q^{\frac{3q-4}{3-2q}} \; (1< q < 3/2)$, $\langle \bar\psi \psi \rangle \propto 1/G_q \; (q<1)$ and $\langle \bar\psi\psi\rangle \propto \log(G_q) / G_q \; (q=1)$. It turns out that in the large coupling limit the chiral condensate tends to zero for $q < 4/3$, while it tends to infinity for $4/3 < q < 3/2$.
The value $q=3/2$ was found to be a limit for a physical system condensate~\cite{Megias:2021jji}. In the critical case $q = 4/3$, the chiral condensate tends to the constant value
 \begin{equation}
     \langle \bar\psi \psi \rangle \stackbin[G_q \to \infty ]{\longrightarrow}{} -\frac{27N_c}{4\pi} \lambda^3 \,.
 \end{equation}

\begin{figure}[t]
	\centering
	\includegraphics[width=0.43\linewidth]{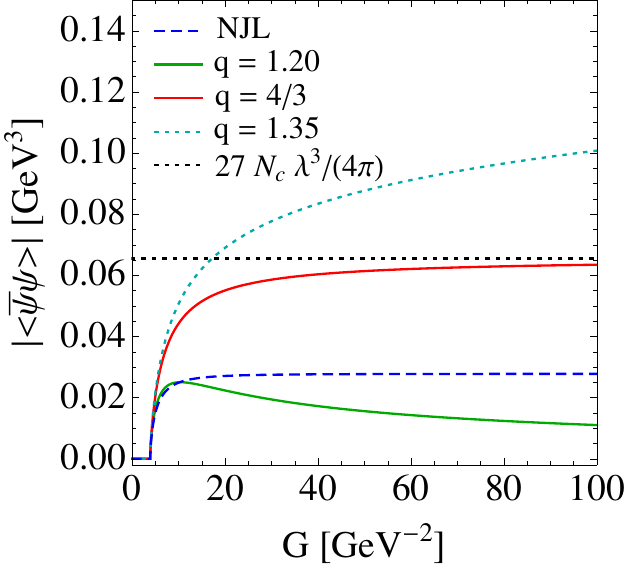} \hspace{0.5cm}
	\includegraphics[width=0.40\linewidth]{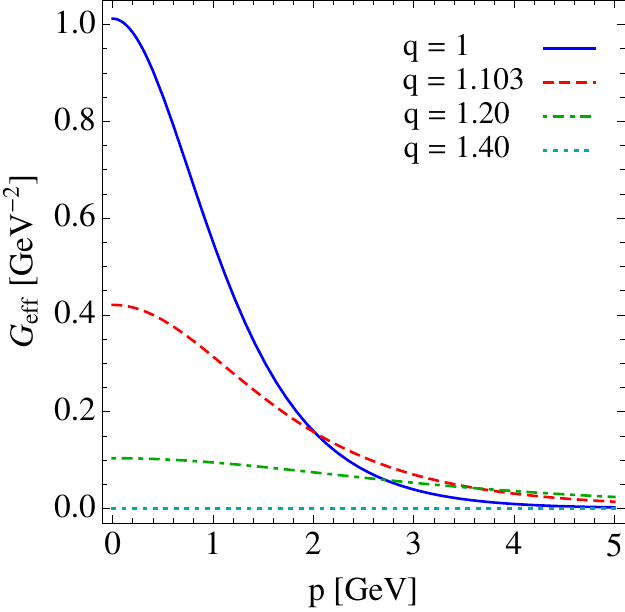}
	\caption{Left panel: Condensate as a function of the coupling up to $G = 100 \,\textrm{GeV}^{-2}$. The horizontal dotted line corresponds to the asymptotic limit of $|\langle \bar\psi \psi\rangle|$ in the \gls{fnjl} model for $q = 4/3$. We have considered $\Lambda = 0.65 \, \textrm{GeV}$ and $m_0 = 0$. The values of $\lambda$ are obtained for each curve by using Eq.~(\ref{eq:lambda_Lambda}). Right panel: Effective two-body coupling as a function of $p$, cf. Eq.~(\ref{eq:Geff}). We display the results for several values of $q$: $q = 1, 1.103, 1.20$ and $1.40$. The dynamical mass, $m$, is obtained for each value of $q$ from the solution of the gap equation, Eq.~(\ref{eq:F}). In this panel, we have used the values $G_q = 5.22 \, \textrm{GeV}^{-2}$, $\lambda = 0.65 \, \textrm{GeV}$and $m_0 = 0$.}
	\label{fig:qq_LargeG}
\end{figure}

The chiral condensate in the NJL model also tends to a constant value in this limit, cf. Eq.~(\ref{eq:qqNJL_largeG}). We display in the left panel of Fig.~\ref{fig:qq_LargeG} the behavior of $|\langle \bar\psi \psi\rangle|$ in the NJL and \gls{fnjl} models up to $G = 100 \,\textrm{GeV}^{-2}$. 
 Note that, in the range $1<q<4/3$, even though the condensate decreases as $G_q$ increases, the product $G_q \langle \bar\psi \psi \rangle$ is still increasing with $G_q$. From Eq.~(\ref{gapequation}), we conclude that the dynamical mass, $m$, also increases with $G_q$. The physical interpretation of this result is the following: for the parameter $q$ in that range, the effective coupling decreases very fast with $p$, then the contribution of the high momentum components of the pair to the effective mass is small, and only less energetic contributions to the self-energy are relevant. This effect is illustrated in the right panel of Fig.~\ref{fig:qq_LargeG}, where the behavior of $G_{\rm eff}(p)$, for several values of the parameter $q$, has been displayed. Thus, the condensate is restricted to quarks with small momenta, and high-momentum bound pairs cannot be formed. The effects of this behaviour of the f\gls{njl} model will be discussed in more detail below. 

\begin{figure}
	\centering
	\includegraphics[width=0.40\linewidth]{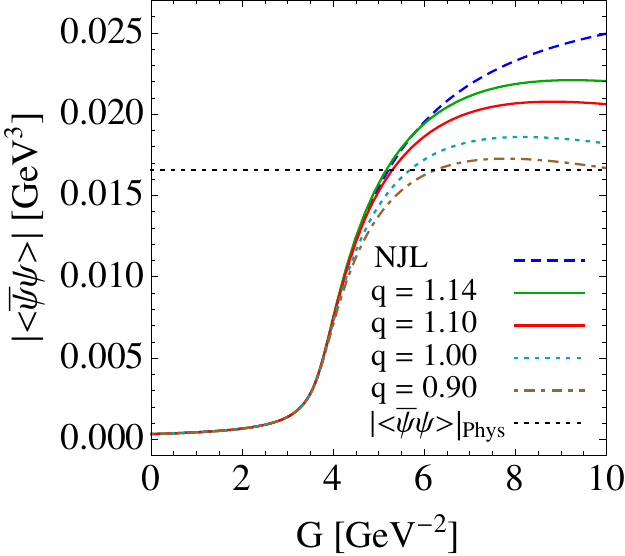} \hspace{0.5cm} 
	\includegraphics[width=0.36\linewidth]{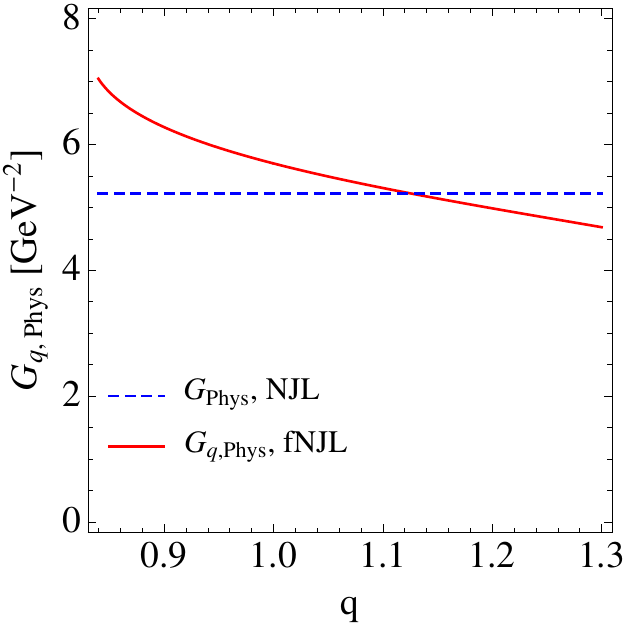}  \hspace{0.5cm}
		\includegraphics[width=0.40\linewidth]{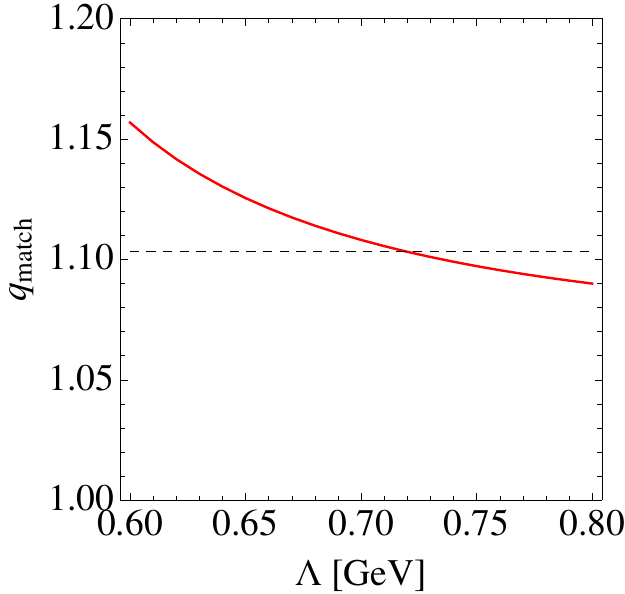} \hspace{0.5cm}
	\includegraphics[width=0.37\linewidth]{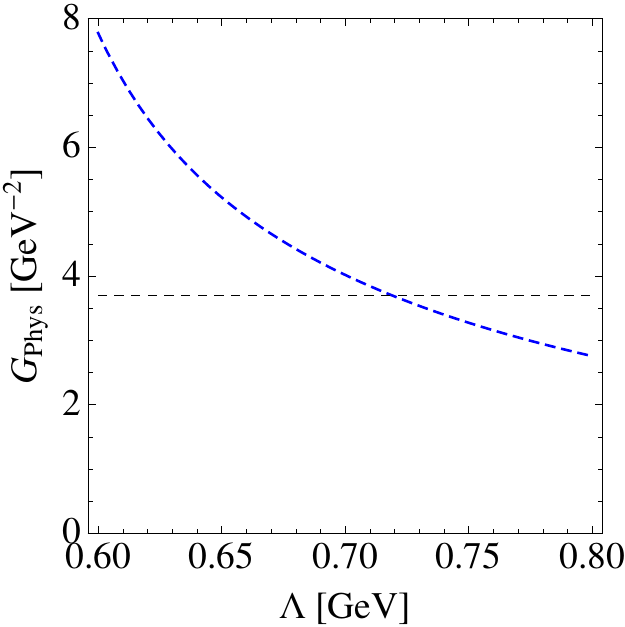}
	\caption{Upper left panel: Condensate as a function of the coupling for $m_0 = 5.0 \, \textrm{MeV}$. The horizontal dotted line corresponds to the physical value of $|\langle \bar\psi \psi \rangle |$ given by Eq.~(\ref{eq:qq_physical}). Upper right panel: Physical value of $G_q$ as a function of $q$ (red line) for $\Lambda = 0.65 \, \textrm{GeV}$. We display also the physical value for the \gls{njl} model, which is $G = 5.22 \, \textrm{GeV}^{-2}$ (dashed blue line). The physical values of $G$ and $G_q$ correspond to the intersection of the horizontal line with each curve in the left panel. The matching between the \gls{njl} and f\gls{njl} results happens at $q_{\textrm{match}} = 1.125$ for $\Lambda = 0.65\, \textrm{GeV}$. Lower left panel: $q_{\textrm{match}}$ as a function of $\Lambda$. To guide the eye we have displayed as an horizontal dashed line the value $q = 1.103$, which is obtained for $\Lambda = 0.719 \, \textrm{GeV}$. Lower right panel: $G_{\textrm{Phys}}$ as a function of $\Lambda$ for the NJL model. The horizontal dashed line corresponds to the value $G_{\textrm{Phys}} = 3.703 \, \textrm{GeV}^{-2}$, which crosses the dashed blue curve at $\Lambda = 0.719 \, \textrm{GeV}$.}
	\label{fig:plotLLqq}
\end{figure}

In Fig.~\ref{fig:plotLLqq} top left panel we show the behaviour of the condensate density obtained through the f\gls{njl} model as a function of $G_q$ (as a function of $G$ for the NJL model) for different values of the parameter $q$, in the regime of the transition between the Wigner-Weyl and Nambu-Goldstone phases. Observe that the critical point for the extended model is practically coincident with that for the standard \gls{njl} model, as all curves almost coincide up to $G_q \sim 4.5 \, \textrm{GeV}^{-2}$ . Above this value, the curves present different behaviours showing a clear dependence on the value for $q$, and they cross the \gls{njl} condensate value at different points. As the value for $q$ increases, the crossing point approaches that of the \gls{njl} model. In Fig.~\ref{fig:plotLLqq} top right panel we see the behaviour of values for $q$ in the f\gls{njl} model that results in the same condensate value obtained by the standard \gls{njl} model, that is, the point where the curves cross the horizontal line in the top left panel of the same figure. We observe that only for $q = 1.125$ the condensate will be the same for the two models for $\Lambda = 0.65 \, \textrm{GeV}$ and for the same values of the couplings, i.e. $G = G_q$. In the following we will denote by $q_{\textrm{match}}$ the value of $q$ the leads to the same values of the couplings, so that $q_{\textrm{match}}(\Lambda)|_{\Lambda = 0.65\,\textrm{GeV}} = 1.125$.

As discussed previously, the fractal model prediction is of $q=1.103$ for $N_f=2$, so we have to change the other parameters in order to obtain the same condensate density in both models when the expected value for $q$ is used, that is, we will fix the value for $q$ and change the other parameters of the model. We have checked that the value for $q_{\textrm{match}}$ is practically independent of $m_0$, so we keep it fixed at $m_0=5.0$~MeV. The dependence of $q_{\textrm{match}}$ with the parameter $\Lambda$, where the relation between $\lambda$ and $\Lambda$ given by Eq.~(\ref{eq:lambda_Lambda}) is considered, is displayed in the bottom left panel of Fig.~\ref{fig:plotLLqq}. From there we get the value $\Lambda = 0.719 \, \textrm{GeV}$ that returns the same value of the condensate for both models in the case $q_{\textrm{match}}=1.103$. Therefore, the f\gls{njl} model is able to reproduce the condensate value given by the \gls{njl} model with the correct value for $q$ calculated according to the fractal approach to \gls{qcd}.

Since we modified the value for $\Lambda$ in the NJL model, we had to recalculate the coupling $G$ that would give the correct physical value for the quark condensate. The behaviour of $G$ with $\Lambda$ in the NJL model is shown in the bottom right panel of Fig.~\ref{fig:plotLLqq}. From this plot we get the value $G_{\textrm{Phys}} = 3.703 \, \textrm{GeV}^{-2}$.

With the method just described, we reproduce the condensate value of the \gls{njl} model by using the f\gls{njl} model with the parameter $q=1.103$ determined by the fractal approach. We summarize the values for the parameters in Table~\ref{table}. In the following subsection we will describe how the pion mass and pion decay constant, which are physical quantities, are determined in the f\gls{njl} model. 

\begin{table}[]
\centering
\begin{tabular}{|l|l|l|}
\hline
parameter & NJL & \gls{fnjl} \\ \hline
\hspace{0.37cm} $\Lambda \;|\; \lambda$ \, \hspace{0.59cm} [GeV] & 0.719  & 0.429    \\ \hline
\hspace{0.34cm} $G \;|\; G_q$  \, \hspace{0.37cm} [GeV$^{-2}$] &  3.703 & 3.703  \\ \hline
$G_{\textrm{crit}} \;|\; G_{q,\textrm{crit}}$  \, [GeV$^{-2}$] &  3.185 & 3.185  \\ \hline
\hspace{0.7cm} $q$ &  \; -  &  1.103  \\ \hline
\end{tabular}
\caption{Table of parameters for the \gls{njl} and f\gls{njl} models that reproduce the same value for the quark condensate $\langle \bar\psi \psi\rangle_{\textrm{Phys}}  = -(0.255\, \textrm{GeV})^3$ with the condition $G_{\textrm{crit}} = G_{q,\textrm{crit}}$, for $m_0 = 5 \, \textrm{MeV}$.}
\label{table}
\end{table}

\subsection{Reproducing the pion mass and the pion decay constant}

We proceed investigating the results of the f\gls{njl} model for the physical parameters, namely, the pions mass and the pions decay constant. 
The pion mass is obtained from the pole of the T-matrix, by the condition~\cite{Vogl:1991qt}
\begin{equation}
    1 - 2G J_\pi(q^2 = m_\pi^2) = 0 \,, \label{eq:2GJ}
\end{equation}
where
\begin{equation}
    J_\pi(q^2) = i \textrm{Tr} \int \frac{d^4p}{(2\pi)^4} \left[ i \gamma_5 \frac{1}{\slash{\!\!\!p} + \frac{\slash{\!\!\!q}}{2} - m + i\epsilon} i \gamma_5 \frac{1}{\slash{\!\!\!p} - \frac{\slash{\!\!\!q}}{2} - m + i \epsilon } \right] \,. 
\end{equation}
In full random phase approximation, the pion mass is obtained as the solution of the following self-consistence equation \cite{Avancini:2016fgq}
 \begin{equation}
    m_\pi^2 = -\frac{m_0}{m} \frac{1}{4iG N_c N_f I(m_\pi^2)} \,,  \label{eq:mpi2}
 \end{equation}
 where $I$ is a function which reads
 \begin{equation}
     I_{\textrm{NJL}}(k^2) =  \frac{i}{8\pi^2} \int_0^1 dx \int_0^\Lambda dp \frac{p^2}{\left( p^2  + \bar m^2(k^2)\right)^{3/2}} = \frac{i}{8\pi^2} \int_0^1 dx \left[ \sinh^{-1}\left(  \frac{\Lambda}{\bar m(k^2)} \right) - \frac{\Lambda}{\sqrt{\Lambda^2 + \bar m^2(k^2)}} \right] \,,
 \end{equation}
 in the \gls{njl} model, and
 \begin{equation}
I_{\textrm{fNJL}}(k^2) =  \frac{i}{8\pi^2} \int_0^1 dx \int_0^\infty dp \frac{p^2}{\left( p^2  + \bar m^2(k^2)\right)^{3/2}}  \left[ 1 + (q-1) \frac{\sqrt{p^2 + \bar m^2(k^2)}}{\lambda} \right]^{-\frac{1}{q-1}} \,, \label{eq:I_qNJL}
 \end{equation}
 in the f\gls{njl} model. In these expressions
 \begin{equation}
\bar m^2(k^2) = m^2 - x(1-x) k^2 \,.
 \end{equation}

The  other observable that can be calculated is the pion decay constant. The quantity is computed in the \gls{njl} model as
\begin{equation}
f^2_{\pi,\textrm{NJL}} = \frac{N_c m^2}{2\pi^2} \int_0^\Lambda dp \frac{p^2}{\left( p^2 + m^2 \right)^{3/2}} = \frac{N_c m^2}{ 2\pi^2 }   \left[ \log\left( x + \sqrt{1+x^2} \right) - \frac{x}{\sqrt{1+x^2}}\right]  \,, \qquad x \equiv \frac{\Lambda}{m} \,,
\end{equation}
while it is computed in the f\gls{njl} model as
\begin{equation}
f^2_{\pi,\textrm{fNJL}} = \frac{N_c m^2}{2\pi^2} \int_0^\infty dp  \frac{p^2}{\left( p^2 + m^2 \right)^{3/2}} \left[ 1 + (q-1) \frac{\sqrt{p^2 + m^2}}{\lambda} \right]^{-\frac{1}{q-1}} \,. \label{eq:fpi_qNJL}
\end{equation}
Notice that the integrands of Eqs.~(\ref{eq:I_qNJL}) and (\ref{eq:fpi_qNJL}) behave as $\propto p^{-1-\frac{1}{q-1}}$ at large momentum, so that the corresponding integrals are convergent in the UV for any value of $q>1$. For the case $q < 1$, the integrations should be performed in the finite interval in momentum provided in Eq.~(\ref{eq:p_interval}), leading thus also to convergent results.

\begin{figure}[b]
 \centering
 	 	\includegraphics[width=7.5cm]{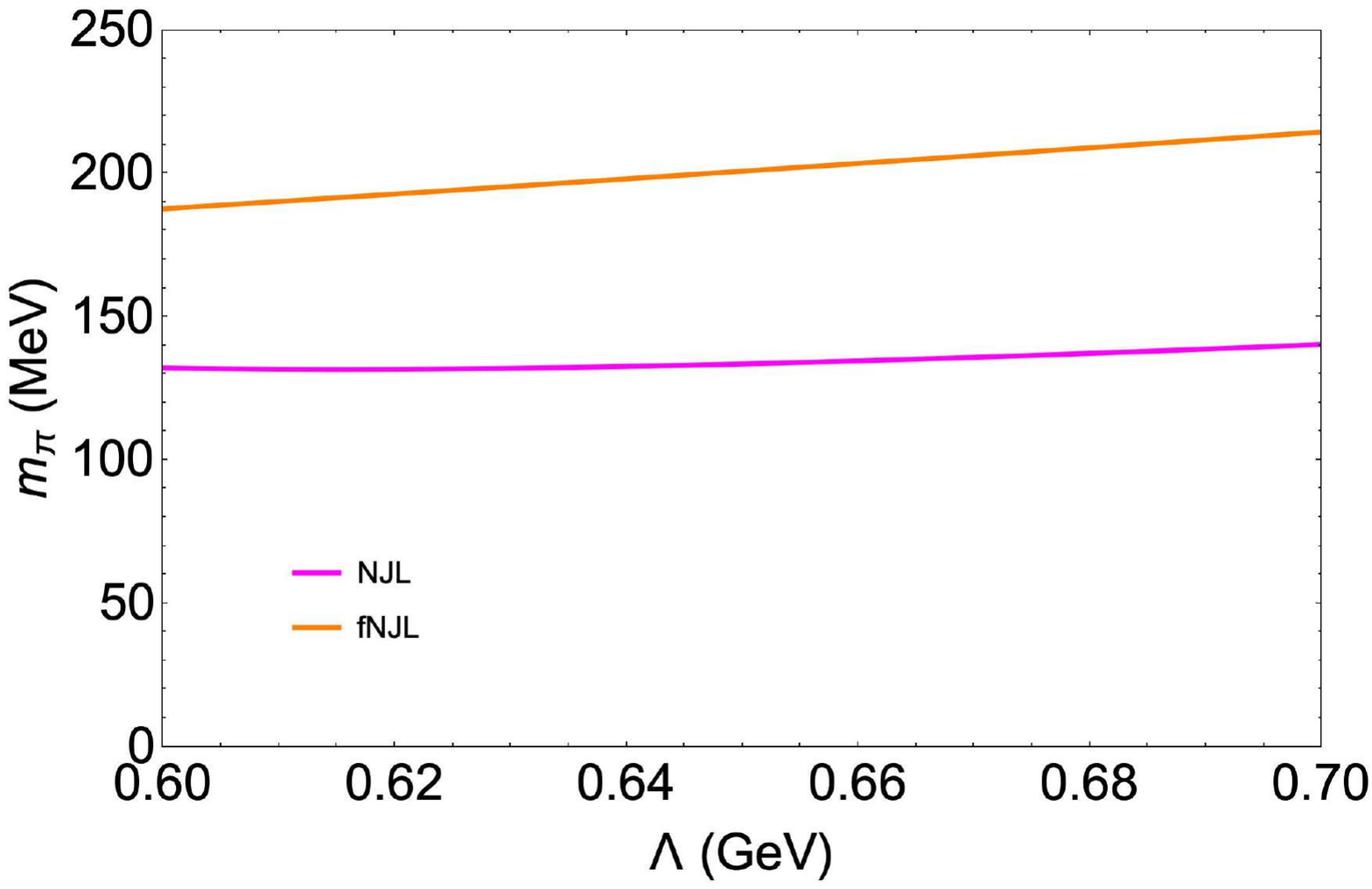} \hspace{0.5cm}	\includegraphics[width=7.5cm]{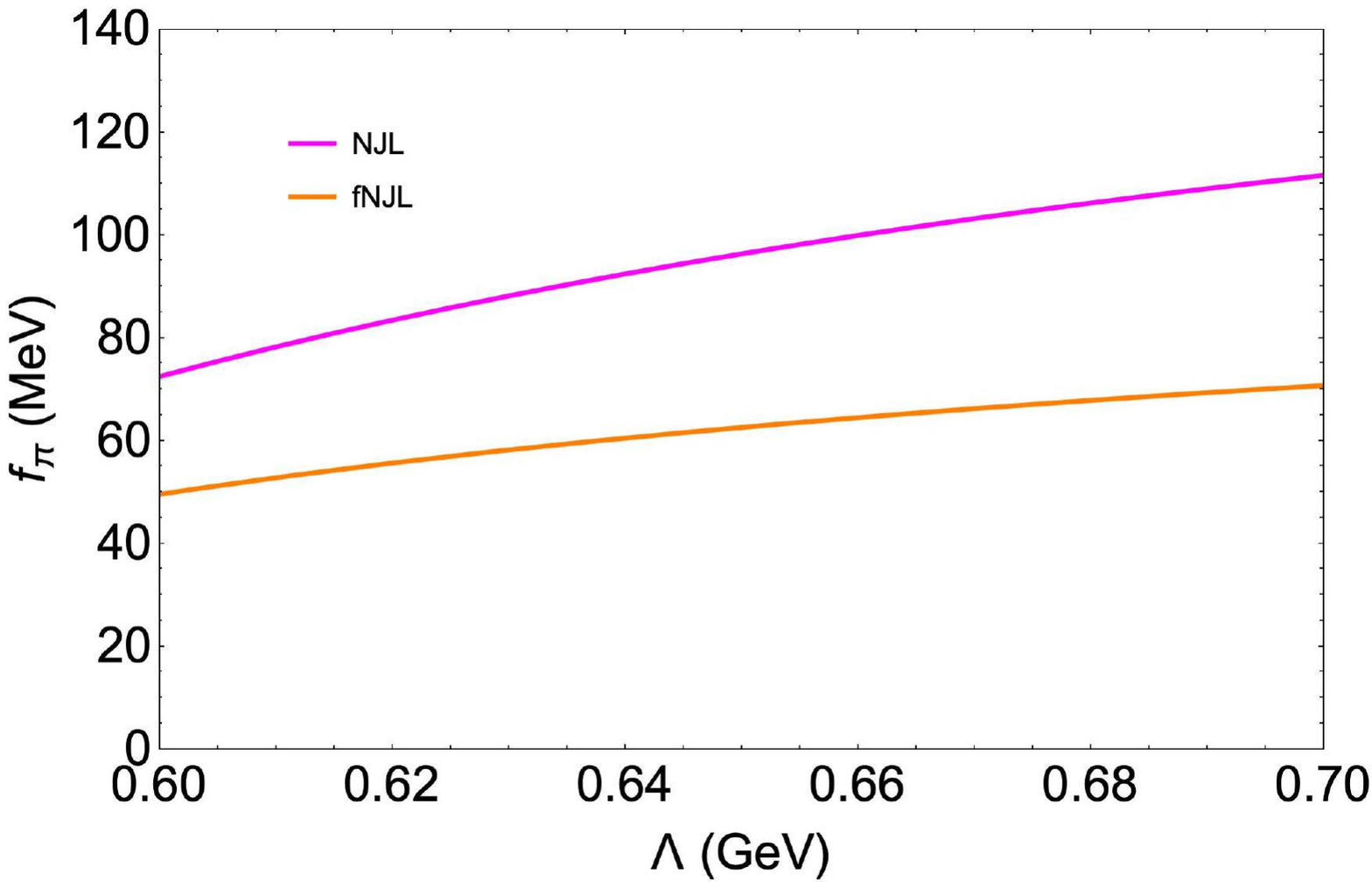} 
	\caption{Left panel: Pion mass $m_\pi$ as a function of the cut-off $\Lambda$ as given by Eq.~(\ref{eq:mpi2}). Right panel: Pion decay constant $f_\pi$ as a function of $\Lambda$. The results were obtained for $G = G_q = 5.22~{\rm GeV}^{-2}$, $q = 1.103$ and $m_0 = 5 \, \textrm{MeV}$, with the relation between $\lambda$ and $\Lambda$ given by Eq.~(\ref{eq:lambda_Lambda}).}
	\label{fig:plot_mpi_fpi}
\end{figure}

We consider the computation of the pion mass and the pion decay constant within the NJL and \gls{fnjl} models in the case where the critical values of the couplings coincide: $G_{\textrm{crit}}$ and $G_{q,\textrm{crit}}$. In this scenario, the relation between $\lambda$ and $\Lambda$ is given by Eq.~(\ref{eq:lambda_Lambda}). Note that the value for the dynamical quark mass, $m$, runs with the cutoff $\Lambda$. So, we maintain the coupling constants fixed, $G = G_q = 5.22 \, \textrm{GeV}^{-2}$, and study the dependence of the observables with the scale $\Lambda$.

In Fig.~\ref{fig:plot_mpi_fpi} we plot the value of the pion mass (left panel) and pion decay constant (right panel) as a function of the parameter $\Lambda$, obtained with the two models. We see that both models give very similar results, presenting comparable cutoff dependencies. At $\Lambda \approx 650~{\rm MeV}$, both models give reasonable values for both the pion mass
and pion decay constant. Naturally, fine tuning is possible. However, the idea here is to show that the fractal inspired coupling leads to similar results as the conventional NJL model with appropriate values for the parameters $q$ and $\lambda$.

%

One of the strengths of the \gls{njl} model is the fact that it satisfies the GOR relation, given by Eq.~(\ref{eq:GOR}). This relation is based on general aspects of the SU(2) symmetry, and thus represents fundamental properties of the hadron structure at low energies. It relates the pion mass and decay constant (hadron side) to the quark bare mass and to the scalar density (quark condensate). Aside from the pion mass and decay constant, we can use the GOR relation to check the quality of the results obtained in our model and if they are similar to the standard NJL model.

\begin{figure}[t]
 \centering
        	 
 	    \includegraphics[width=7.5cm]{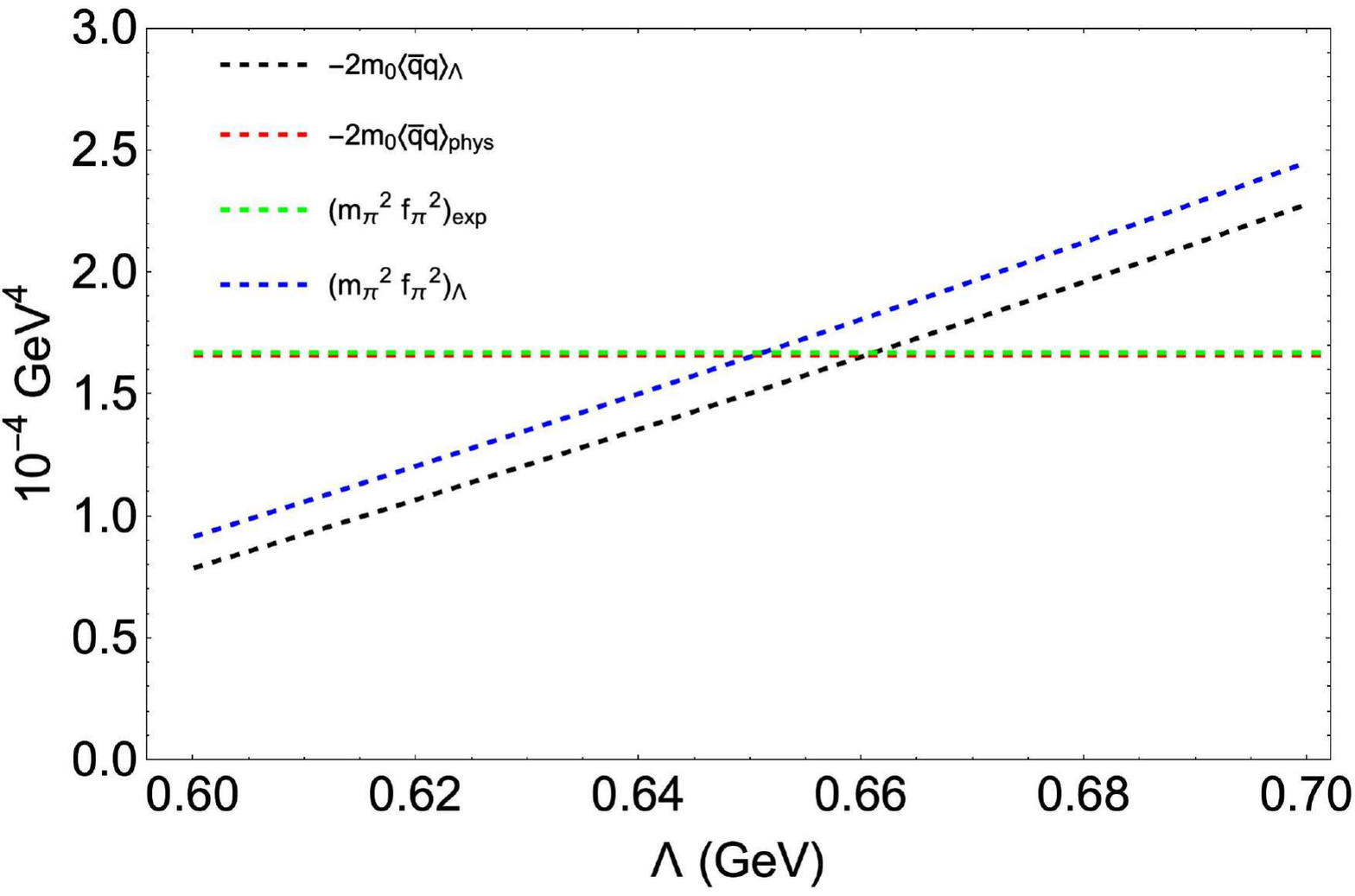} 
 	    \hspace*{0.5cm}
 	    \includegraphics[width=7.5cm]{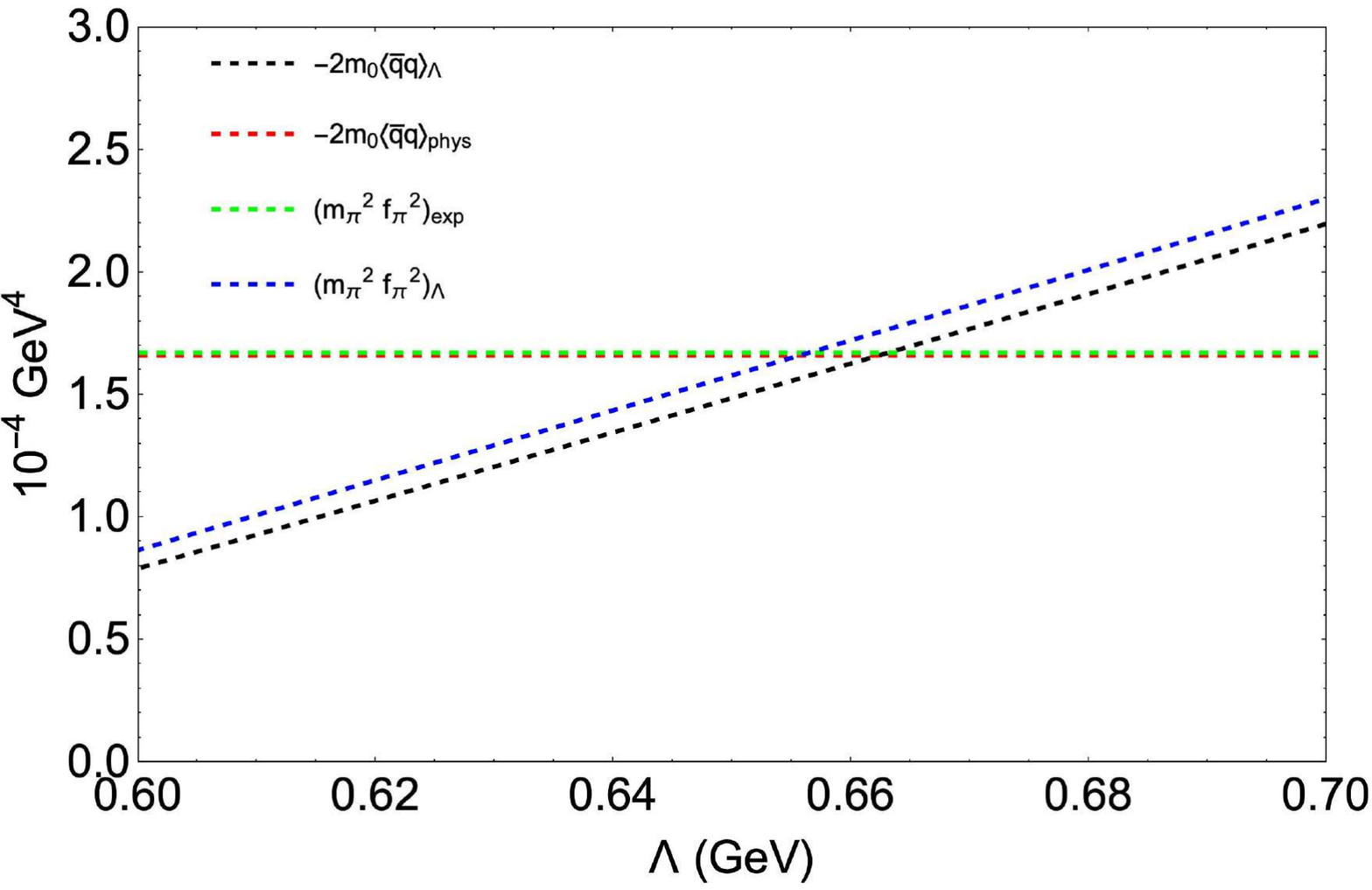}
 	\caption{Check of the \gls{gor} relation for several cutoffs in the NJL (left) and
 	fNJL (right) models. The results were obtained for $G = G_q = 5.22~{\rm GeV}^{-2}$, $q = 1.103$ and $m_0 = 5 \, \textrm{MeV}$, with the relation between $\lambda$ and $\Lambda$ given by Eq.~(\ref{eq:lambda_Lambda}).}
 	\label{fig:GMOR}
 \end{figure}

In Fig.~\ref{fig:GMOR}, we display a GOR relation check obtained by computing 
$-2m_0\times\langle \bar q q \rangle$ and $m^2_\pi \times f^2_\pi$ in both models compared to physical and experimental values. The left panel shows the cutoff dependence in the NJL model and the right panel shows the cutoff dependence in the \gls{fnjl} model. We can observe that both models give the same $\Lambda$-dependence and they match the physical and experimental values at very close scales.



\section{Conclusions and outlook}

In this work we have studied the generalization of the NJL model by the inclusion of an effective coupling inspired in the fractal structure of the Yang-Mills fields. The running coupling constant works as a natural regularization method for the theory, eliminating the divergences that are present in the standard NJL model. More explicitly, we have provided in the present work a physical explanation, in terms of a fractal QCD vacuum, for a running coupling that renormalizes the quark condensate. So, even though the effective coupling resembles just another non-local NJL model regulator, here it is natural outcome
of the fractal structure of the QCD vacuum.

We studied the fractal model in the scenario where the critical coupling is the same in both NJL and \gls{fnjl} models and obtained the pion mass and the pion decay constant. We show that the \gls{gor} relation is satisfied for similar cutoffs in both models.

The physical scenario for the f\gls{njl} model results in a different behavior of the condensate, resulting in a different critical coupling with respect to the NJL model. Also, the physical coupling, that reproduces the pion properties, is smaller than that of the \gls{njl} model, and closer to the critical value than what is observed in the NJL model.

Since the running coupling constant used is inspired in the fractal approach used in \gls{hep} to study the multi particle production in high energy collisions, the generalization of the \gls{njl} model presented here allows to connect the hadron structure properties to the properties of the hot and dense matter obtained in those collisions. This model opens the possibility to study, in a unified way, the \gls{qcd} properties in the high energy limit and in the low energy limit of the hadron structure. This unified procedure counts with the consistent fractal approach to the \gls{qcd} that results in the fractal-inspired effective coupling. Let us stress that this work is focused on the physics of the fNJL model at zero temperature. It would be interesting to extend it to finite temperature. Outcomes similar to those in Refs.~\cite{Rozynek:2015zca,Zhao:2020xob,Mitra:2017cib,Islam:2023zpl} are expected.

We have seen that in the range $1<q<4/3$, an interesting phenomenon takes place: while the condensate density decreases, the hadron mas still increases with the coupling strength. By using Eqs.~(\ref{eq:q_Nc_Nf}) and (\ref{eq:qq_asymptotic}), one can see that the behavior of $\langle \bar\psi \psi\rangle \stackbin[G_q \to \infty ]{\longrightarrow}{} 0$ in the \gls{fnjl} model, which corresponds to $q < 4/3$, is obtained~for
\begin{equation}
    N_f < \frac{1}{2} \left( 11 N_c - 9 \right) \,, \label{nf}
\end{equation}
while values of $N_f$ larger than that lead to $\langle \bar\psi \psi\rangle \stackbin[G_q \to \infty ]{\longrightarrow}{} \infty$. Then, for $N_c = 3$ a vanishing value of the chiral condensate in the large coupling limit is obtained for $N_f < 12$, while in the case $N_c = 1 = N_f$ the \gls{fnjl} model would be in the critical situation $q = 4/3$. 

For any non Abelian field theory with $N_f$ in agreement with the relation~(\ref{nf}), only the pairs quark/anti-quark with low momenta would condensate, but the hadron mass would, nevertheless, increase. In a field theory with the structure parameters in the range above, we would have massive partons but no condensate for very large couplings. In this case, we can conjecture that there would be no formation of a system similar to the QGP, and the high energy processes would result in the direct production of particles. The coalescence mechanism would be less important, and the particle multiplicity would be higher, and with more massive particles.

This mechanism might be the responsible for production of dark-matter, in particular it could constitute a mechanism of dark-matter formation based on the sector of the non-Abelian Yang-Mills theory.
It would be interesting also to perform a more detailed analysis on the $N_c$ and $N_f$ dependence of the chiral condensate, with applications on the phase diagram of QCD, see e.g. Ref.~\cite{Ahmad:2022hbu} for a recent research in this line.

The model introduced in this work opens the possibility to investigate new aspects of the hadronic matter near the transition between the Wigner-Weyl and the Nambu-Goldstone phases. In particular, the role of the different flavours and their masses, or the effective number of flavours in the formation of the condensate. The model allows to study how the critical value, or its existence, depends on the number of colours and flavours. Also, we may include thermal and magnetic effects in our model and study
the thermodynamics and the phase diagram with the fractal effective coupling.


\section*{Acknowledgments}
The work of E M is supported by the project PID2020-114767GB-I00 and by the Ram\'on y Cajal program under Grant RYC-2016-20678 funded by MCIN/AEI/10.13039/501100011033 and by ``FSE Investing in your future'', by the FEDER/Junta de Andaluc\'{\i}a-Consejer\'{\i}a de Econom\'{\i}a y Conocimiento 2014-2020 Operational Program under Grant A-FQM-178-UGR18, by Junta de Andaluc\'ia under Grant FQM-225, by the Consejer\'{\i}a de Conoci\-miento, Investigaci\'on y Universidad of the Junta de Andaluc\'{\i}a and European Regional Development Fund (ERDF) under Grant SOMM17/6105/UGR, and by the ``Pr\'orrogas de Contratos Ram\'on y Cajal'' Program of the University of Granada.
M J T is supported by S\~ao Paulo Research Foundation (FAPESP) grant 2021/12954-5.
V S T is supported by FAPESP grant 2019/010889-1 and by the Conselho Nacional de Desenvolvimento Cient\'{\i}fico e Tecnol\'ogico (CNPq) grant 305004/2022-0. 
A D is partially supported by the CNPq, grant 304244/2018-0, by Project INCT-FNA Proc. No. 464 898/2014-5, and by FAPESP grant 2016/17612-7. 


\linespread{.9}
\bibliographystyle{ieeetr}
\bibliography{paper.bib}

\end{document}